\documentclass[vecphys]{svmult}

\usepackage{makeidx}         
\usepackage{graphicx}        
\usepackage{multicol}        
\usepackage{cite}            
\usepackage[bottom]{footmisc}
\usepackage{amsmath,amssymb}

\newcommand{\mi}{\mathrm{i}}
\newcommand{\half}{\textstyle{\frac{1}{2}}}
\newcommand{\threehalves}{\textstyle{\frac{3}{2}}}

\makeindex             

\begin{document}

\title{Mean-field description of multicomponent exciton-polariton superfluids}
\author{Y.~G.~Rubo}
\institute{Centro de Investigaci\'on en Energ\'{\i}a, Universidad Nacional
 Aut\'onoma de M\'exico, Temixco, Morelos, 62580, Mexico
 \texttt{ygr@cie.unam.mx}}

\maketitle

\begin{abstract}
This is a review of spin-dependent (polarization) properties of multicomponent
exciton-polariton condensates in conditions when quasi-equilibrium mean-field
Gross-Pitaevskii description can be applied. Mainly two-component (spin states
$\pm1$) polariton condensates are addressed, but some properties of
four-component exciton condensates, having both the bright (spin $\pm1$) and
the dark (spin $\pm2$) components, are discussed. Change of polarization state
of the condensate and phase transitions in applied Zeeman field are described.
The properties of fractional vortices are given, in particular, I present
recent results on the warping of the field around half-vortices in the presence
of longitudinal-transverse splitting of bare polariton bands, and discuss the
geometrical features of warped half-vortices (in the framework of the lemon,
monstar, and star classification).
\end{abstract}

\section{The Gross-Pitaevskii equation}
 \label{sec:GPE}
 \index{Gross-Pitaevskii equation}
Currently, most theoretical descriptions of exciton-polariton condensates
observed \cite{Kasprzak06,Balili07,Lai07,Baumberg08,Wertz09} in incoherently
excited semiconductor microcavities are based on the Gross-Pitaevskii equation
(GPE). When the polarization of the condensate is of interest, this equation
can be generically written as
\begin{equation}
 \label{GPEgener}
 \mi\hbar\frac{\partial}{\partial t}\vec{\psi}(\vec{r},t)
 =\frac{\delta H}{\delta\vec{\psi}^*(\vec{r},t)}, \qquad
 H=\int \mathcal{H}(\vec{\psi}^*,\vec{\psi})d^2r,
\end{equation}
where the order parameter $\vec{\psi}(\vec{r},t)$ of the condensate is a
complex 2D vector function of the 2D position in the microcavity plane
$\vec{r}$ and time $t$. Alternatively, one can expand $\vec{\psi}$ on the
circular polarization basis
\begin{equation}
  \label{psicirc}
 \vec{\psi}
 =\frac{\hat{\vec{x}}+\mi\hat{\vec{y}}}{\sqrt{2}}\psi_{+1}
 +\frac{\hat{\vec{x}}-\mi\hat{\vec{y}}}{\sqrt{2}}\psi_{-1},
\end{equation}
and obtain the coupled GPEs for two circular components $\psi_{\pm1}$ (see
Eqs.~(\ref{GM_GPEdetailed}a,b) in Section \ref{sec:LeMonStar}).

GPE \eqref{GPEgener} is used in two main flavors, strongly non-equilibrium GPE
and quasi-equilibrium GPE, that treat the energy relaxation in two extreme
ways. It is completely neglected in the former, and it is considered to be
essential in the later. Mathematically, these approaches differ in Hamiltonian
density $\mathcal{H}(\vec{\psi},\vec{\psi}^*)$: it is complex in the former and
it is real in the later. Each approach has its benefits and drawbacks.

The imaginary part of the Hamiltonian for non-equilibrium GPE
\cite{Wouters07a,Wouters07b,Keeling08a,Borgh10} is given by the difference of
income and escape rates of exciton-polaritons into and out of the condensate.
While the escape rate is given by the reciprocal radiative life-time of
exciton-polaritons and is independent of the particle density, the income rate
is a non-linear function of it. The nonlinearity is essential to stabilize the
solution and it appears due to the depletion of an incoherently pumped
reservoir. This approach resulted to be quite successful in modeling the
experimental data on condensate density profiles for spatially nonuniform
exciton-polariton condensates. On the other hand, it cannot describe the
spontaneous formation of linear polarization of the condensate---the fact that
is quite unfortunate since the observation of spontaneous linear polarization
is one of the direct experimental evidences of Bose-Einstein condensation of
exciton-polaritons \cite{Kasprzak06,Balili07}. A workaround is to add the
Landau-Khalatnikov relaxation into \eqref{GPEgener}, which is equivalent to
adding an imaginary part to time $t$. But this relaxation is rather artificial
because it changes the number of particles in the condensate.

In what follows, we consider the opposite limit assuming a fast relaxation of
exciton-polaritons, so that they reach quasi-equilibrium, even with the
temperature that can be different from the lattice one. The balance of income
and outcome rates produces some steady-state concentration of
exciton-polaritons, that can be defined, as usual, by introducing the chemical
potential $\mu$. %
\index{Chemical potential} The coherent fraction of condensed particles can be
described by the traditional GPE with real
$\mathcal{H}(\vec{\psi},\vec{\psi}^*)$,
\begin{equation}
 \label{Ham}
 \mathcal{H}=\mathcal{T}-\mu n+\mathcal{H}_\mathrm{int}+\mathcal{H}^\prime.
\end{equation}
Here $\mathcal{T}$ is the density of the kinetic energy,
$\mathcal{H}_\mathrm{int}$ describes interaction between the particles,
$\mathcal{H}^\prime$ stands for some possible perturbations, and
$n=\vec{\psi}^*\cdot\vec{\psi}$ is the exciton-polariton density.

The kinetic energy of exciton-polaritons in planar microcavities depends on the
orientation of vector $\vec{\psi}$ with respect to the direction of motion.
Near the bottom of lower polariton branch one has
\begin{subequations}
  \label{KinEnergy}
\begin{eqnarray}
 \mathcal{T} &=& \frac{\hbar^2}{2m_l}|\vec{\nabla}\cdot\vec{\psi}|^2 +
                 \frac{\hbar^2}{2m_t}|\vec{\nabla}\times\vec{\psi}|^2
 \\ &=&
 \frac{\hbar^2}{m_l}
 \left|\frac{\partial\psi_{+1}}{\partial z^*}+\frac{\partial\psi_{-1}}{\partial z}\right|^2
 +
 \frac{\hbar^2}{m_t}
 \left|\frac{\partial\psi_{+1}}{\partial z^*}-\frac{\partial\psi_{-1}}{\partial z}\right|^2,
\end{eqnarray}
\end{subequations}
where $m_l$ and $m_t$ are the longitudinal and transverse effective masses
\index{TE-TM splitting}
of polaritons, and the complex derivatives
\begin{equation}
 \label{ComplexDeriv}
 \frac{\partial}{\partial z}=\frac{1}{2}\left(
 \frac{\partial}{\partial x}-\mi\frac{\partial}{\partial y}
 \right), \qquad
 \frac{\partial}{\partial z^*}=\frac{1}{2}\left(
 \frac{\partial}{\partial x}+\mi\frac{\partial}{\partial y}
 \right),
\end{equation}
are used. The vector $\vec{\psi}$ is proportional to the in-plane electric
field vector of exciton-polariton mode. According to \eqref{KinEnergy}, the
frequency of transverse electric (TE) mode with in-plane wave vector
$\vec{k}\perp\vec{\psi}$ is $\hbar k^2/2m_t$, while for the transverse magnetic
(TM) mode with $\vec{k}\|\vec{\psi}$ the frequency is $\hbar k^2/2m_l$. (The
same bare frequencies of both modes at $k=0$ are removed from
\eqref{KinEnergy}.)

The polariton-polariton interaction is also anisotropic: it depends on mutual
orientation of $\vec{\psi}$ and $\vec{\psi}^*$. One can construct two quartic
invariants from these two vectors and $\mathcal{H}_\mathrm{int}$ is given by
\begin{subequations}
\label{Hinter}
\begin{eqnarray}
 \mathcal{H}_\mathrm{int}
 &=&\frac{1}{2}(U_0-U_1)(\vec{\psi}^*\!\cdot\vec{\psi})^2
 +\frac{1}{2}U_1|\vec{\psi}^*\times\vec{\psi}|^2 \\
 &=&\frac{1}{2}U_0\left(|\psi_{+1}|^4+|\psi_{-1}|^4\right)
 +(U_0-2U_1)|\psi_{+1}|^2|\psi_{-1}|^2.
\end{eqnarray}
\end{subequations}
It is seen that $U_0$ is the amplitude of interaction of polaritons with the
same circular polarization (with the same spin), and $U_0-2U_1$ is the
amplitude of interaction of polaritons with opposite circular polarizations
(opposite spins). These quantities are denoted by $\alpha_1$ and $\alpha_2$ in
some papers. The constant $U_0$ is positive and can be estimated as
$\sim\mathcal{E}_ba_B^2$, where $\mathcal{E}_b$ is the exciton binding energy
and $a_B$ is the exciton Bohr radius. The interaction of exciton-polaritons
with opposite spins depends substantially on the electron-electron and
hole-hole exchange processes and is defined by the electron and hole
confinement within quantum wells and by the number of quantum wells in the
microcavity. As a result, the value of $U_1$ is sensitive to the microcavity
geometry.

To end this section it is important to mention the limitations of any GPE in
application to the condensates of exciton-polaritons in microcavities, or to
condensates of any other bosonic excitations that have a finite radiative
life-time. Due to interference of light emitted from different parts of
condensate there appears dissipative long-range coupling in the system. Most
importantly, the escape rate becomes dependent on the symmetry of the
condensate wave-function and this favors the formation of particular
long-living many-particle states \cite{Aleiner11}. These effects cannot be
properly treated in the framework of Gross-Pitaevskii equation
\eqref{GPEgener}.

\section{Polarization and effects of Zeeman field}
 \label{sec:Zeeman}
 \index{Zeeman field}
The interaction energy \eqref{Hinter} of the polariton condensate is
polarization dependent. While the first term in (\ref{Hinter}a) does not depend
on polarization and is simply proportional to the square of the polariton
concentration $n=(\vec{\psi}^*\!\cdot\vec{\psi})$, the second term in
(\ref{Hinter}a) is sensitive to the degree of the circular polarization of the
condensate. For $U_1>0$ the interaction energy is minimized when the second
term in (\ref{Hinter}a) is annulated, which is achieved for polarization
satisfying $\vec{\psi}^*\times\vec{\psi}=0$, i.e., for the linear polarization.
On the other hand, in the case $U_1<0$ the minimum is reached for the circular
polarization of the condensate, when $\vec{\psi}^*\times\vec{\psi}=\pm\mi n$.

So, there is qualitative change in the ground state of the condensate when
$U_1$ changes sign \cite{Shelykh06}.
\begin{enumerate}
 \item[(i)]{$U_1>0$. The ground state is characterized by two angles, the
     total phase angle $\theta$ and the polarization angle $\eta$. These
     angles are defined from the Descartes components of the order
     parameter $\psi_x=\sqrt{n}\,e^{\mi\theta}\cos\eta$ and
     $\psi_y=\sqrt{n}\,e^{\mi\theta}\sin\eta$. The circular components are
     then $\psi_{\pm1}=\sqrt{n/2}\,e^{\mi(\theta\mp\eta)}$. There are two
     broken continuous symmetries and, consequently, the excitation
     spectrum consists of two Bogoliubov branches. The sound velocities for
     these branches at $m_l=m_t=m^*$ are $v_0=\sqrt{\mu/m^*}$ and
     $v_1=\sqrt{nU_1/m^*}$, where $\mu=(U_0-U_1)n$ is the chemical
     potential. The presence of TE-TM splitting leads to the anisotropy of
     sound velocities (see \cite{Shelykh06} for details).}
     \vspace{0.5\baselineskip}
 \item[(ii)]{$U_1<0$. In this case one of the circular components is zero
     and the other is $\sqrt{n}\,e^{\mi\theta}$. Since there is only one
     broken continuous symmetry, the excitation spectrum consists of only
     one Bogoliubov branch, and the other branch is gaped parabolic with
     the gap $2|U_1|n$. The chemical potential is $\mu=U_0n$ in this
     domain, so that the sound velocity for the Bogoliubov excitations is
     $\sqrt{\mu/m^*}$.}
\end{enumerate}
\index{Bogoliubov dispersion} \index{Chemical potential}

The mean-field theory predicts an arbitrary polarization for $U_1=0$ since in
this case the energy of the condensate is polarization independent. In reality,
fluctuations destroy the order in this case at any finite temperature $T$. It
can be already understood from the excitation spectrum, because, apart from the
Bogoliubov branch, there is the gapless parabolic branch with dispersion
$\hbar^2k^2/2m^*$ and the condensate would evaporate completely due to
excitation of these quasiparticles.\footnote{The concentration of
quasiparticles with the energy $\epsilon(k)$ is given by $\int (2\pi)^{-2}
[\exp\{\epsilon(k)/T\}-1]^{-1}d^2k$ and the integral diverges logarithmically
for small $k$ when $\epsilon(k)\propto k^2$.} One can also map this case to the
O(4) nonlinear sigma model, \index{Nonlinear sigma model} where the order is
proven to be absent for $T>0$ \cite{Pelissetto02}.

Note the similarity between the two-component condensates of exciton-polaritons
and three-component condensates of spin-1 cold atoms \cite{Ho98,Ohmi98}. Due to
the 3D rotational symmetry, there are also only two interaction constants in
the latter case. These constants are defined by the cross-sections of
scattering of two atoms with the total spin 0 and 2.\footnote{The case of the
total spin 1 is irrelevant since the orbital wave function of colliding bosons
is antisymmetric and it cannot be realized within the condensate.} Two
different atomic condensates can also be found depending on the sign of the
scattering length with the total spin 2:  ferromagnetic and anti-ferromagnetic
(or polar), which are analogs of circularly and linearly polarized
exciton-polariton condensates, respectively.

It is the first case, $U_1>0$, that is realized in the exciton-polariton
condensates observed so far. The linearly polarized condensate can be seen as
composition of equal numbers mutually coherent spin-up and spin-down
polaritons. Therefore, it is interesting to study the effect of applied
magnetic field to this state \cite{Rubo06}. Considering only weak fields, when
the magnetic length is much greater than the exciton Bohr radius, one can study
only the effects of Zeeman field, that is described by adding
\begin{equation}
 \label{ZF_Def}
 \mathcal{H}^\prime=\Omega\,(|\psi_{-1}|^2-|\psi_{+1}|^2)
\end{equation}
into Hamiltonian \eqref{Ham}. Here the Zeeman field $\Omega$ is given
by the half of the Zeeman splitting energy for a single polariton.

To find the order parameter for the uniform condensate in the presence of
Zeeman field it is convenient to introduce the concentrations of the components
$n_{\pm1}=|\psi_{\pm1}|^2$ satisfying $n_{+1}+n_{-1}=n$. Assuming both $n_{+1}$
and $n_{-1}$ to be nonzero, one can take variations of the Hamiltonian
\begin{multline}
 \label{ZF_UniformH}
 \mathcal{H}_\mathrm{int}+\mathcal{H}^\prime-\mu n \\
 =\frac{1}{2}U_0\left(n_{+1}+n_{-1}\right)^2
 -2U_1n_{+1}n_{-1}-(\mu+\Omega)n_{+1}-(\mu-\Omega)n_{-1}
\end{multline}
over $n_{\pm1}$ to obtain
\begin{equation}
 \label{ZF_TwoEqs}
 -2U_1n_{\pm1}=(\mu-U_0n\mp\Omega).
\end{equation}
The sum and the difference of Eqs.\ \eqref{ZF_TwoEqs} results in
\begin{equation}
 \label{ZF_SubCrit}
 \mu=(U_0-U_1)n, \quad
 n_{\pm1}=\frac{1}{2}\left(n\pm\frac{\Omega}{U_1}\right), \quad
 \mathrm{for}\;|\Omega|<\Omega_c\equiv nU_1.
\end{equation}
For higher Zeeman fields, $|\Omega|>\Omega_c$, one of the components becomes
empty, $n_{-1}$ for $\Omega>\Omega_c$ and $n_{+1}$ for $\Omega<-\Omega_c$, and
in this case $\mu=U_0n-|\Omega|$.

Remarkably, for subcritical fields the chemical potential does not change at
all, so that there is no change in the position of the emission line. %
\index{Chemical potential} The only effect of applied Zeeman field is the
change of circular polarization degree
$\varrho_c=(n_{+1}-n_{-1})/(n_{+1}+n_{-1})=\Omega/nU_1$, that increases
linearly with the field. The elliptical polarization of the condensate for
subcritical fields is characterized by two angles, and in the the same way as
for the linearly polarized condensate, there are two Goldstone modes; only the
sound velocities change with the Zeeman field. This implies the full
suppression of the Zeeman splitting by polariton-polariton interactions within
the condensate \cite{Rubo06}. Note also that for subcritical fields there are
two phase transitions in the left and in the right circular component of the
condensate, respectively \cite{Keeling08b}. The Zeeman splitting (the gap in
the exciton spectrum) appears only for supercritical fields $|\Omega|>\Omega_c$
where the condensate becomes circularly polarized. This effect, observed
experimentally by Larionov \emph{et al.} \cite{Larionov10}, allows to measure
the spin-dependent interaction constant $U_1$. \index{Zeeman splitting}

\section{Vortices in exciton-polariton condensates}
 \label{sec:Vortices}
Vortices play a key role in various physical phenomena both on macroscopic and
microscopic level. While the vortex formation is very important for description
of different effects in fluid mechanics, in particular, in aerodynamics and
turbulent flow motion, the understanding of properties of quantized vortices is
crucial for description of phase transitions in condensed matter. The well
known examples are the phase transitions in type II superconductors in applied
magnetic field, which are related to the formation and melting for vortex
lattices\cite{Abrikos57}, and the Berezinskii-Kosterlitz-Thouless (BKT) phase
transition\cite{Berez72,KT73,Koster74}. \index{Berezinskii-Kosterlitz-Thouless
transition}

As it was discussed above, the exciton-polariton condensates possess
two-component order parameter \eqref{psicirc} and these condensates allow
half-quantum vortices (half-vortices) \cite{Rubo07}. Moreover, the
half-vortices are basic topological excitations in this case (see
\cite{VolovikBook} for a review on basic properties of half-quantum vortices).
In spite of recent observation of both integer \cite{Lagoudakis08} and
half-integer \cite{Lagoudakis09} vortices in exciton-polariton condensates, the
presence of half-vortices was recently questioned \cite{Flayac10} for the case
of two-band dispersion with TE-TM splitting of polariton band given by Eq.\
\eqref{KinEnergy}. In this section I present the details on how the vortex
solutions should be found in this case (a short summary of this theory has been
given in \cite{Miller10a}). In what follows only the case of zero Zeeman field
will be considered. \index{Half-vortex}

For a 2D system of radius $R$ the energy of a vortex is finite but
logarithmically large,
\begin{equation}
 \label{VortexEng}
 E_\mathrm{vor}=E_c+E_s\ln(R/a),
\end{equation}
where $a=\hbar/\sqrt{2m^*\mu}$ is the characteristic radius of vortex core (the
effective mass $m^*$ is defined below in Eq.~\eqref{gamma}). The fact that
$E_\mathrm{vor}$ diverges logarithmically at $R\rightarrow\infty$ is good: it
prevents the single vortices to be excited at low temperatures and thus
protects the long-range order of the condensate. Knowledge of prefactor $E_s$
allows to estimate \cite{Koster74} the BKT transition temperature $T_c$. The
proliferation of single vortices appears when the free energy
$E_\mathrm{vor}-TS$ crosses zero. The vortex core area is $a^2$ and it can be
appear in $R^2/a^2$ places, so that the entropy $S=\ln(R/a)^2$ and this gives
$T_c=E_s/2$ if one neglects the energy of the core $E_c$ in \eqref{VortexEng}.
The energy $E_s\ln(R/a)$ is elaborated on large distances from the vortex core
$r\gg a$, which we will refer to as the elastic region, and the study of
vortices should begin with establishing the behavior of the order parameter in
this region. When TE-TM splitting is present this behavior is, in general,
nontrivial.

\subsection{The order parameter on large distances}
In the elastic region the order parameter changes within the order parameter
manyfold, i.e., the polarization of the condensate is linear everywhere in this
domain. The circular-polarization components $\psi_{\pm1}$ defined in
\eqref{psicirc} can be written in cylindrical coordinates $(r,\phi)$ as
\begin{equation}
 \label{Asymp}
 \psi_{\pm1}(r\gg a,\phi)=\sqrt{\frac{n}{2}}\,e^{\mi[\theta(\phi)\mp\eta(\phi)]},
\end{equation}
where $n$ is the constant concentration of the condensate at large distances,
and the phases are written in terms of total phase angle $\theta$ and
polarization angle $\eta$. These angles do not depend on the radius $r$ (such
dependence would only increase the vortex energy), but they are functions of
the azimuthal angle $\phi$. Since the order parameters should be uniquely
defined in the whole space, one has
\begin{equation}
 \label{WindNumbers}
 \eta(\phi+2\pi)-\eta(\phi)=2\pi k, \qquad
 \theta(\phi+2\pi)-\theta(\phi)=2\pi m.
\end{equation}

These conditions divide all possible solutions of GPE into topological
sectors. Each sector is defined by two topological charges (or winding
numbers), $k$ and $m$. The state from one sector cannot be continuously
transformed into another sector, or, in other words, any state of the
condensate evolves within its own topological sector. The sector
$k=m=0$ is the ground state sector; the minimum energy here is reached
for position-independent order parameter. By definition, the vortex is
the state that minimizes the energy in a topological sector with at
least one non-zero winding number. The energy of the $(k,m)$-vortex
\eqref{VortexEng} is counted from the ground state energy, i.e., it is
the difference between the minimal energy in the $(k,m)$ sector and the
minimal energy of $(0,0)$ sector (the ground state energy). Since only
the sum and the difference, $\theta\pm\eta$, enter Eq.\ \eqref{Asymp},
the winding numbers can be either both integer or both half-integer,
and the corresponding vortices are referred accordingly. Note also that
the vortex corresponds to a minimum of Hamiltonian $H$ for specific
boundary conditions: $\delta H/\delta\vec{\psi}^*=0$ for the vortex
solution and, therefore, it is a static solution of GPE
\eqref{GPEgener}.\footnote{Note, however, that this does not imply that
a single vortex gives an absolute minimum of the $H$ in the
corresponding topological sector. For example, the integer vortex
$(1,0)$ can be unstable with respect to decay into the pair of
$(\half,\half)$ and $(\half,-\half)$ half-vortices for $m_l<m_t$ (see
subsection \ref{sec:Vortices}.2 below).}

According to \eqref{WindNumbers} one can add any periodic functions of
$\phi$ to $\eta(\phi)$ and $\theta(\phi)$ without changing the
topological sector. The proper functions $\eta(\phi)$ and
$\theta(\phi)$ for the $(k,m)$-vortex should be found from minimization
of Hamiltonian in elastic region. The corresponding part of Hamiltonian
is related solely to the kinetic energy term $\int\mathcal{T}d^2r$.
After substitution of \eqref{Asymp} into (\ref{KinEnergy}b) and use of
the asymptotic behavior of the complex derivative for
$r\rightarrow\infty$,
\begin{equation}
 \label{DDzLargeR}
 \frac{\partial}{\partial z}\rightarrow
 -\frac{\mi}{2r}e^{-\mi\phi}\frac{\partial}{\partial\phi},
\end{equation}
one obtains the product of integrals over $r$ and $\phi$ that results in the
second term of Eq.\ \eqref{VortexEng}. The integral over $r$ diverges
logarithmically and should be cut by the core size $a$ at small $r$, and by
system radius $R$ at large $r$. This gives the factor $\ln(R/a)$. The prefactor
is then given by
\begin{equation}
 \label{Prefactor1}
 E_s=\frac{\hbar^2n}{2m^*}\int_0^{2\pi}
 \left\{
 [1+\gamma\cos(2u)](1+u^\prime)^2+[1-\gamma\cos(2u)]\theta^{\prime2}
 \right\}d\phi,
\end{equation}
where the prime denotes the derivative over $\phi$ and
\begin{equation}
 \label{Eta&u}
 u(\phi)=\eta(\phi)-\phi.
\end{equation}
The effective mass $m^*$ and the TE-TM splitting parameter $\gamma$ are
defined in \eqref{Prefactor1} by
\begin{equation}
 \label{gamma}
 \frac{1}{m^*}=\frac{1}{2}\left(\frac{1}{m_l}+\frac{1}{m_t}\right), \qquad
 \gamma=\frac{m_t-m_l}{m_t+m_l}.
\end{equation}

Variations of the functional \eqref{Prefactor1} over $\theta$ and $u$
lead to the equations
\begin{subequations}
 \label{Eqs_phi}
 \begin{equation}
 \left[1-\gamma\cos(2u)\right]\theta^{\prime\prime}+
 2\gamma\sin(2u)u^{\prime}\theta^{\prime}=0,
 \end{equation}
 \begin{equation}
 \left[1+\gamma\cos(2u)\right]u^{\prime\prime}
 +\gamma\sin(2u)\left( 1-u^{\prime2} -\theta^{\prime2} \right)=0.
 \end{equation}
\end{subequations}
In general, the polarization will be radial at least at one specific direction
and it is convenient to count the azimuthal angle from this direction and set
the total phase to be zero at this direction as well. Then, the solutions of
Eqs.~(\ref{Eqs_phi}a,b) for $(k,m)$-vortex should satisfy the boundary
conditions
\begin{subequations}
 \label{BoundCond}
\begin{equation}
 u(0)=0, \qquad \theta(0)=0,
\end{equation}
\begin{equation}
u(2\pi)=2(k-1)\pi, \qquad \theta(2\pi)=2m\pi.
\end{equation}
\end{subequations}

The solutions in question are trivial for some particular vortices.
\begin{enumerate}
 \item[(i)]{\emph{Hedgehog vortices.} These are $(1,m)$-vortices having
     $\theta=m\phi$ and $u\equiv 0$, so that the polarization angle
     $\eta=\phi$. Polarization points into the radial direction everywhere
     and these vortices look like hedgehogs. These solutions are similar to
     magnetic monopoles \cite{Rajaraman}. } \index{Hedgehog vortex}
     \index{Monopole} \vspace{0.5\baselineskip}
 \item[(ii)]{\emph{Double-quantized polarization vortex $(2,0)$.} In this
     special case $\theta\equiv 0$, but $u=\phi$, resulting in
     $\eta=2\phi$. Polarization rotates twice when one encircles the vortex
     core. These vortices and experimental possibilities of their
     excitation in exciton-polaritons fields have been studied by Liew
     \emph{at el.} \cite{Liew07}. }
\end{enumerate}

\begin{figure}[t]
\centering
\includegraphics*[width=0.9\textwidth]{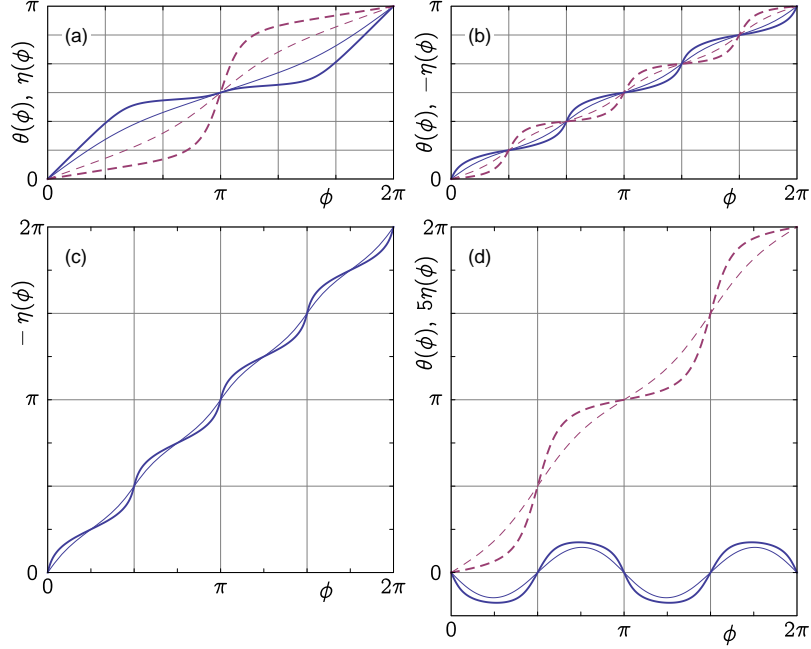}
\caption[]{The dependence of polarization angle $\eta$ (solid lines) and
 phase angle $\theta$ (dashed lines) on the azimuthal angle $\phi$ for
 two values of TE-TM splitting parameter:
 $\gamma=-0.4$ (thin lines) and $\gamma=-0.9$ (thick lines).
 The panels show the behavior of the angles for
 the $(\half,\half)$ half-vortex (a), the $(-\half,\half)$ half-vortex (b),
 the $(-1,0)$ polarization vortex (c), and the $(0,1)$ phase vortex (d).
 In the last case the periodic function $\eta(\phi)$ has been
 upscaled for clarity}
\label{Fig1-Angles}
\end{figure}

In other cases the solutions should be found numerically. Both
Eqs.~(\ref{Eqs_phi}a) and (\ref{Eqs_phi}b) can be integrated once to give
\begin{subequations}
 \label{Eqs_phi1}
 \begin{equation}
 \left[1-\gamma\cos(2u)\right]\theta^\prime=C_1,
 \end{equation}
 \begin{equation}
 \left[1+\gamma\cos(2u)\right]u^{\prime2}+
 \left[1-\gamma\cos(2u)\right]\theta^{\prime2}
 -\gamma\cos(2u)=C_2,
 \end{equation}
\end{subequations}
and the solutions can be written as integrals of elementary functions. The
constants $C_{1,2}$ should then be found, e.g., by shooting, to satisfy the
boundary conditions (\ref{BoundCond}). The functions $\theta(\phi)$ and
$\eta(\phi)$ are shown in Fig.~\ref{Fig1-Angles} for elementary half-vortices
and for two integer vortices $(-1,0)$ and $(0,1)$, that also exhibit nonlinear
dependencies of polarization and phase angles.

Fig.~\ref{Fig1-Angles} demonstrates the behavior of angles for negative values
of the TE-TM splitting parameter $\gamma$. The functions $\theta(\phi)$ and
$\eta(\phi)$ for positive $\gamma$ can be found by the shift. Indeed, the
change $u\rightarrow u+(\pi/2)$ in Eqs.~(\ref{Eqs_phi}a,b) results in the
change of the sign of $\gamma$. More precisely, to satisfy the boundary
conditions \eqref{BoundCond} the transformations can be written as
\begin{subequations}
\begin{eqnarray}
 \label{TransSign}
 && \gamma\rightarrow -\gamma, \\
 && u(\phi)\rightarrow u\left(\phi+\frac{\pi}{2(k-1)}\right)+\frac{\pi}{2}, \\
 && \theta(\phi)\rightarrow
 \left(\phi+\frac{\pi}{2(k-1)}\right)-\theta\left(\frac{\pi}{2(k-1)}\right),
\end{eqnarray}
\end{subequations}
and they can be applied to all vortices except the hedgehogs with $k=1$ (and
where they are not necessary, of course, since $u(\phi)\equiv0$).

The nonlinear change of angles seen in Fig.~\ref{Fig1-Angles} becomes
especially evident when $\gamma$ approaches $\pm1$. This limit
correspond to a strong inequality between effective masses, e.g.,
$m_l\gg m_t$ for $\gamma\rightarrow -1$. Qualitatively the strong
nonlinearities can be understood if one introduces the effective masses
for the phase $m_\theta$ and for the polarization $m_\eta$,
\begin{equation}
 \label{AngleMasses}
 \frac{1}{m_\theta}=\frac{\cos^2\!u}{m_t}+\frac{\sin^2\!u}{m_l}, \qquad
   \frac{1}{m_\eta}=\frac{\sin^2\!u}{m_t}+\frac{\cos^2\!u}{m_l},
\end{equation}
and writes the energy \eqref{Prefactor1} as
\begin{equation}
 \label{Prefactor2}
 E_s=\frac{\hbar^2n}{2}\int_0^{2\pi}\left\{
 \frac{\eta^{\prime2}}{m_\eta}+\frac{\theta^{\prime2}}{m_\theta}
 \right\} d\phi.
\end{equation}
The effective masses \eqref{AngleMasses} depend on the orientation of
polarization. Since $u(\phi)$ changes between the values specified by
\eqref{BoundCond}, there are sectors where $m_\theta\approx m_l$ and
$m_\eta\approx m_t$, and there are sectors where $m_\theta\approx m_t$
and $m_\eta\approx m_l$. To minimize the energy \eqref{Prefactor2} in
the case $m_l\gg m_t$, the phase angle changes rapidly and the
polarization angle stays approximately constant in the former, while
there is the opposite behavior in the latter.

\subsection{The energies and interactions of vortices}
In the absence of TE-TM splitting the energies $E_s$ of vortices are
\begin{equation}
 \label{Es_gamma0}
 E_{s0}^{(k,m)}=E_0(k^2+m^2), \qquad E_0=\frac{\pi\hbar^2n}{m^*},
 \qquad (\mathrm{for}\;\gamma=0).
\end{equation}
It is seen that in this case the energy of an elementary half-vortex is exactly
half of the energy of an elementary integer vortex. Important consequences can
be drawn from this relation concerning the interactions between half-vortices.
The four elementary half-vortices can be divided in two kinds, right
half-vortices with $k+m=\pm1$, and left ones with $k-m=\pm1$. One can see from
\eqref{Asymp} the the right half-vortices possess the vorticity of the
left-circular component of the order parameter, the amplitude of this component
goes to zero and the phase of this component becomes singular in the vortex
core center, and, as a result, the polarization becomes right-circular in the
core center. For left half-vortices the picture is opposite. It follows from
\eqref{Es_gamma0} that the left and the right half-vortices do not interact
with each other. Consider, for example, the $(\half,\half)$ and
$(-\half,\half)$ half-vortices. The elastic energy of this pair is
$E_0\ln(R/a)$ both when they are far away from each other and when they are in
the same place forming the phase vortex $(0,1)$. So, the only possible coupling
between the left and the right half-vortices is of short range, related to the
overlap of their cores and resulting change of the core energy term $E_c$ in
\eqref{VortexEng}. So, the only long-range coupling is present between the
half-vortices of the same kind. It can be shown that identical half-vortices
repel each other logarithmically, while the half-vortices and
anti-half-vortices, $(k,m)$ and $(-k,-m)$, attract each other logarithmically,
as it is in the case of vortices and antivortices in one-component condensates
\cite{ChaikinLub}. This simple picture is changed in the presence of TE-TM
splitting that leads to the long-range interaction between half-vortices of
different kind.

\begin{figure}[t]
\centering
\includegraphics*[width=0.7\textwidth]{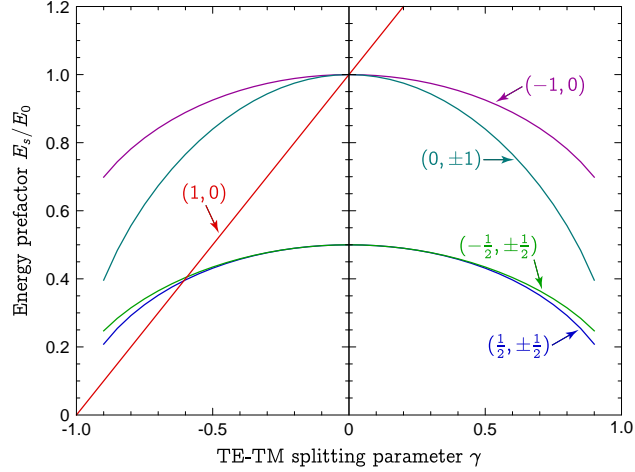}
\caption[]{The logarithmic prefactor of vortex energies $E_s$
(see Eqs.\ \eqref{VortexEng} and \eqref{Prefactor1})
for half-vortices and integer vortices as functions of TE-TM
splitting parameter $\gamma$ \eqref{gamma}.
The curves are labeled by the winding numbers $(k,m)$ of the vortices,
and the energies are given in the units of $E_0=\pi\hbar^2n/m^*$}
\label{Fig2-Es}
\end{figure}

The logarithmic prefactors $E_s^{(k,m)}$ for elementary half-vortices and
elementary integer vortices are shown in Fig.~\ref{Fig2-Es}. For all of them,
expect the hedgehog (1,0)-vortex, these energies are even functions of
$\gamma$, which can be proven using the transformations \eqref{TransSign}.
These energies decrease with increasing $\gamma^2$. The case of hedgehog is
special, as it has been discussed above. The hedgehog polarization is radial
everywhere and $E_s^{(1,0)}$ is defined purely by the longitudinal effective
mass,
\begin{equation}
 \label{Es_HedgV}
 E_s^{(1,0)}=\pi\hbar^2n/m_l=E_0(1+\gamma).
\end{equation}
It can be shown that when two vortices with winding numbers $(k_1,m_1)$ and
$(k_2,m_2)$ are injected in the condensate and they are separated by distance
$r$, such that $a\ll r\ll R$, the energy of the condensate is increased in
logarithmic approximation (i.e., omitting the core energies) by
\begin{multline}
 \label{InterGen}
 E_s^{(k_1+k_2,m_1+m_2)}\ln(R/a) \\
 +\left[E_s^{(k_1,m_1)}+E_s^{(k_2,m_2)}-E_s^{(k_1+k_2,m_1+m_2)}\right]\ln(r/a).
\end{multline}
The second term in \eqref{InterGen} gives the interaction energy of two
vortices. The coupling between vortices arising due to TE-TM splitting
can be analyzed analytically in the limit of small $\gamma$.

The solutions of Eqs.~(\ref{Eqs_phi}a,b) for $k\ne1$ are written as series in
$\gamma$,
\begin{subequations}
 \label{SeriesAngles}
\begin{multline}
 \theta(\phi)=m\phi +\frac{m}{2(k-1)}\,\gamma\sin[2(k-1)\phi] \\
 +\frac{m[(k-1)^2-(1-m^2)]}{16(k-1)^3}\,\gamma^2\sin[4(k-1)\phi]+\dots,
\end{multline}
\begin{multline}
 u(\phi)=(k-1)\phi-\frac{[(k-1)^2+(m^2-1)]}{4(k-1)^2}\,\gamma\sin[2(k-1)\phi] \\
 +\frac{[5(k-1)^4+2(k-1)^2(m^2-3)+(m^2-1)^2]}{64(k-1)^4}
 \,\gamma^2\sin[4(k-1)\phi]+\dots.
\end{multline}
\end{subequations}
Substitution of these expression into \eqref{Prefactor1} gives
\begin{equation}
 \label{SeriesEs}
 \frac{E_s^{(k,m)}}{E_0}=(k^2+m^2)
 -\frac{[k^2(k-2)^2+2[2+3k(k-2)]m^2+m^4]}{8(k-1)^2}\,\gamma^2-\dots,
\end{equation}
and, in particular,
\begin{equation}
 \label{Es_forIntgV}
 E_s^{(-1,0)}=E_0\left[1-\frac{9}{32}\gamma^2-\dots\right], \qquad
 E_s^{(0,\pm1)}=E_0\left[1-\frac{5}{8}\gamma^2-\dots\right].
\end{equation}
There is no difference between the energies of half-vortices at this order of
$\gamma$-series. The difference, however, appears in the next order. The series
for the angles up to $\gamma^4$ are rather cumbersome to be presented, but they
result in
\begin{subequations}
 \label{Es_forHV}
\begin{equation}
 E_s^{\left(\half,\pm\half\right)}=
 \frac{E_0}{2}\left[1-\frac{\gamma^2}{2}-\frac{3\gamma^4}{16}-\dots\right],
\end{equation}
\begin{equation}
 E_s^{\left(-\half,\pm\half\right)}=
 \frac{E_0}{2}\left[1-\frac{\gamma^2}{2}-\frac{11\gamma^4}{144}-\dots\right].
\end{equation}
\end{subequations}

Eqs.\ \eqref{Es_HedgV}, \eqref{Es_forIntgV}, and \eqref{Es_forHV} can be used
to find the interactions between half-vortices according to Eq.\
\eqref{InterGen}. Most important interaction that appears due to TE-TM
splitting is between the $(\half,\half)$ and $(\half,-\half)$ half-vortices.
For small $\gamma$ their coupling constant is linear in $\gamma$ and the
interaction energy is \cite{Miller10b}
\begin{equation}
 \label{LRCoupling}
 V_{(\half,\half),(\half,-\half)}\simeq -\gamma E_0\ln(r/a).
\end{equation}
It should be noted that the interrelation between $m_t$ and $m_l$, i.e., the
sign of $\gamma$, depends on the detuning of the frequency of the cavity photon
mode from the center of the stop-band of the distributed Bragg mirror
\cite{Panzarini99}. So, one can have both attraction and repulsion of the
$(\half,\half)$ and $(\half,-\half)$ half-vortices. The coupling of the other
left and right half-vortices is quadratic in $\gamma$. The $(-\half,\half)$ and
$(-\half,-\half)$ half-vortices repel each other with the interaction energy
being $-(7/32)\gamma^2E_0\ln(r/a)$. The $(-\half,\pm\half)$ and
$(\half,\pm\half)$ half-vortices attract each other with the interaction energy
being $(1/8)\gamma^2E_0\ln(r/a)$.

In the absence of TE-TM splitting there is no coupling between the right
half-vortices (with $km>0$) and the left ones (with $km<0$) and there are two
decoupled BKT transitions, corresponding to the dissociation of pairs of left
and right half-vortices \cite{Rubo07,Keeling08b}. The transition temperature is
then estimated from the energy of single half-vortex as $E_0/4$. The TE-TM
splitting of polariton bands changes this picture substantially. First, because
all four half-vortices become coupled and, secondly, because the energies of a
vortex and its antivortex become different, so it is not clear with one should
be used in the estimation of critical temperature.

One expects qualitative modifications of the BKT transition in the region of
$\gamma$ close to $-1$. In this region the attraction of the $(\half,\half)$
and $(\half,-\half)$ half-vortices becomes very strong and, as a result, the
hedgehog is the vortex with the smallest energy in the system for
$\gamma<\gamma_c\simeq -0.6$ (see Fig.~\ref{Fig2-Es}). It does not mean,
however, that the transition temperature can be estimated from the energy of
the hedgehog in this region. In fact, the phase transition occurs due to
dissociation of vortex-antivortex pairs, and the energy of the $(1,0)$ and
$(-1,0)$ pair is still bigger than the energy of the pair of two half-vortices.
One expects that when pairs of half-vortices are thermally excited in the
system they will tend to form molecules consisting of the hedgehog (formed by
merging of the $(\half,\half)$ and $(\half,-\half)$ half-vortices) with the
$(-\half,-\half)$ and $(-\half,\half)$ half-vortices being attached to it. The
proliferation of these $(-\half,-\half)-(1,0)-(-\half,\half)$ molecules defines
the phase transition for $m_l\gg m_t$. \index{Berezinskii-Kosterlitz-Thouless
transition}

\section{Geometry of the half-vortex fields}
 \label{sec:LeMonStar}
In general, two coupled Gross-Pitaevskii equations for the circular components
of the order parameter \index{Gross-Pitaevskii equation}
\begin{subequations}
 \label{GM_GPEdetailed}
 \begin{multline}
 \mi\hbar\frac{\partial\psi_{+1}}{\partial t}
 =-\frac{\hbar^2}{2m^*}\left(
 \Delta\psi_{+1}+4\gamma\frac{\partial^2\psi_{-1}}{\partial z^2}
 \right)-\mu\psi_{+1} \\
 +\left[U_0(|\psi_{+1}|^2+|\psi_{-1}|^2)-2U_1|\psi_{-1}|^2\right]\psi_{+1},
 \end{multline}
\begin{multline}
 \mi\hbar\frac{\partial\psi_{-1}}{\partial t}
 =-\frac{\hbar^2}{2m^*}\left(
 \Delta\psi_{-1}+4\gamma\frac{\partial^2\psi_{+1}}{\partial z^{*2}}
 \right)-\mu\psi_{-1} \\
 +\left[U_0(|\psi_{-1}|^2+|\psi_{+1}|^2)-2U_1|\psi_{+1}|^2\right]\psi_{-1},
 \end{multline}
\end{subequations}
are not separated in cylindrical coordinates $(r,\phi)$. The variables are
separated only in special cases of hedgehog vortices and the double-quantized
polarization vortex discussed in previous section after Eqs.~\eqref{BoundCond}.
For other vortices one needs to solve GPEs numerically in two spacial
dimensions with the boundary conditions at large distances defined by Eqs.\
\eqref{Asymp} and \eqref{Eqs_phi}.

\begin{figure}[t]
\centering
\includegraphics*[width=0.9\textwidth]{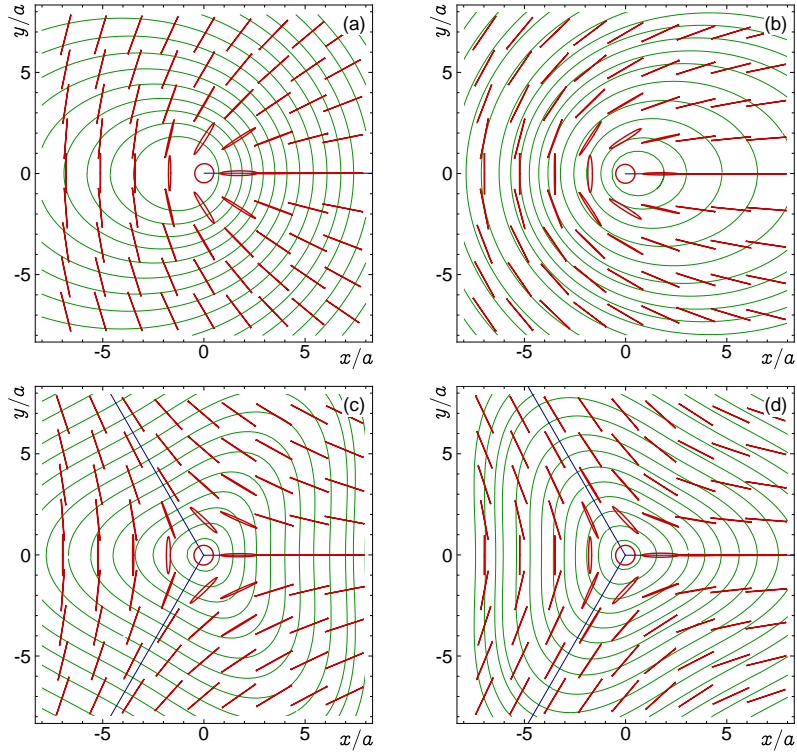}
\caption[]{Showing the geometry of half-vortices for different values of
 TE-TM splitting parameter $\gamma$ \eqref{gamma}. The plots are obtained
 from numerical solutions of GPEs \eqref{GM_GPEdetailed} \cite{Miller12}.
 The interaction constants are related as $U_1=0.55U_0$.
 The local polarization ellipses are drawn with the thick (red) lines.
 The streamlines of the current are shown by thin (green) lines.
 The panels demonstrate the following cases:
 (a) the half-vortices $(\half,\pm\half)$ for $\gamma=-0.5$ (the lemon morphology);
 (b) the half-vortices $(\half,\pm\half)$ for $\gamma=0.5$ (the lemon morphology);
 (c) the half-vortices $(-\half,\pm\half)$ for $\gamma=-0.5$ (the star morphology);
 (d) the half-vortices $(-\half,\pm\half)$ for $\gamma=0.5$ (the star morphology).
 }
\label{Fig3-LemonStar}
\end{figure}

Each vortex with nonzero phase winding number $m$ is characterized by a finite
superfluid current circulating around its core center. Performing numerical
solutions it is important to take into account the fact that streamlines of the
current are deformed with respect to perfect circles in the presence of TE-TM
splitting. The circular components of the current $\vec{J}$ are given by
\begin{multline}
 \label{GM_Jplus}
 J_{+1}=\frac{\mi\hbar}{\sqrt{2}\,m^*}\left\{\left(
  \psi_{+1}\frac{\partial\psi_{+1}^*}{\partial z}
 +\psi_{-1}\frac{\partial\psi_{-1}^*}{\partial z}
 -\psi_{+1}^*\frac{\partial\psi_{+1}}{\partial z}
 -\psi_{-1}^*\frac{\partial\psi_{-1}}{\partial z}
 \right)\right. \\
 +2\gamma\left.\left(
  \psi_{+1}\frac{\partial\psi_{-1}^*}{\partial z^*}
 -\psi_{-1}^*\frac{\partial\psi_{+1}}{\partial z^*}
 \right)\right\},
\end{multline}
and $J_{-1}=J_{+1}^*$. They are related to the radial $J_r$ and the azimuthal
$J_\phi$ components of the current by
\begin{equation}
 \label{GM_Jr&Jphi}
 J_r=\frac{1}{\sqrt{2}}\left(
 e^{\mi\phi}J_{+1}+e^{-\mi\phi}J_{-1} \right), \qquad
 J_\phi=\frac{\mi}{\sqrt{2}}\left(
 e^{\mi\phi}J_{+1}-e^{-\mi\phi}J_{-1} \right).
\end{equation}
At large distances $r\gg a=\hbar/\sqrt{2m^*\mu}$ one can use Eq.~\eqref{Asymp}
to find
\begin{subequations}
 \label{Currents}
 \begin{equation}
 J_r=\frac{\hbar n}{m^*r}\gamma\sin(2u)\frac{d\theta}{d\phi},
 \end{equation}
 \begin{equation}
 J_\phi=\frac{\hbar n}{m^*r}\left[1-\gamma\cos(2u)\right]\frac{d\theta}{d\phi}.
 \end{equation}
\end{subequations}
Note that the condition of conservation of the total number of polaritons for
the static vortex solution of Eqs.~\eqref{GM_GPEdetailed},
\begin{equation}
 \label{GM_divJ}
 \mathrm{div}\vec{J}=\frac{1}{r}\left[
 \frac{\partial}{\partial r}(rJ_r)+\frac{\partial J_\phi}{\partial\phi}
 \right]=0,
\end{equation}
implies $\partial J_\phi/\partial\phi=0$. So, Eq.~(\ref{Eqs_phi}a) obtained in
the previous section is in fact the condition of conservation of the azimuthal
current.

The warping of streamlines of current is shown in Fig.~\ref{Fig3-LemonStar}.
The order parameter has been found numerically \cite{Miller12} for different
values of TE-TM splitting parameter. To find the static solutions of GPE
\eqref{GM_GPEdetailed}, we have been choosing an initial order parameter
$\vec{\psi}(\vec{r},t=0)$ satisfying the boundary conditions that follow from
Eqs.~\eqref{Eqs_phi} and that are shown in Fig.~\ref{Fig1-Angles} for a given
topological sector $(k,m)$. Apart from this the initial functions were
continuous but arbitrary. Then the functions were evolved according to
\eqref{GM_GPEdetailed} in imaginary time. As a result, the order parameter
relaxed to corresponding static half-vortex solution. The resulting half-vortex
solutions are found to be independent of the initial shape of
$\vec{\psi}(\vec{r},t=0)$.

In Fig.~\ref{Fig3-LemonStar} one can see two distinct morphologies of basic
half-vortices. The geometry of half-vortex solutions can be discussed in terms
of singular optics \cite{Nye99,Soskin01}, where the polarization singularity
related to a half-vortex is referred as C-point, to indicate that the
polarization is circular at the vortex center and, therefore, the direction of
the main axis of polarization ellipse is not defined. The morphologies of the
field around C-points are classified by the index of associated real tensor
field, and, additionally, by the number of strait polarization
lines\footnote{The tangents of polarization lines define by the direction of
the main axis of polarization ellipse in each point.} that terminate at C-point
\cite{Dennis09}. The tensor index coincides with the polarization winding
number $k$, and the number of lines could be either one or three. As a result,
three different morphologies can be found. Following Berry and Hannay
\cite{Berry77,Nye83}, these morphologies are referred as \emph{lemon},
\emph{star}, and \emph{monstar}. \index{Lemon, star, monstar}

The lemon configuration is characterized by $k=\half$ and by only one straight
polarization line terminating in the vortex center. This is the morphology of
vortices in Fig.~\ref{Fig3-LemonStar}a,b with the straight polarization line
being defined by $\phi=0$. The star configuration is characterized by
$k=-\half$. In this case there are always three straight lines terminating in
the vortex center. The stars are realized in Fig.~\ref{Fig3-LemonStar}c,d and
three straight polarization lines are defined by $\phi=0,\pm2\pi/3$.

The change of parameter $\gamma$ leads to deformation of polarization texture
and to deformation of streamlines of the current, but it does not result in the
change of morphologies of half-vortices. In principle, one could expect the
transformation of lemon into monstar, since these morphologies possess the same
topological index $k=\half$. Contrary to the lemon case, however, the monstar
is characterized by three straight polarization lines terminating in the vortex
center, similar to the star configuration. So, from geometrical point of view
the monstar has got intermediate structure between the lemon and the star, and
this is why its name is constructed from ``(le)mon-star''.

To have the monstar configuration one needs a special behavior of polarization
angle $\eta(\phi)$. Namely, it is necessary to have
\begin{equation}
 \label{GM_MonstarCond}
 \left.\frac{d\eta}{d\phi}\right|_{\phi=0}>1 \quad
 \mathrm{for}\;k=\frac{1}{2}.
\end{equation}
In this case the polarization angle initially rotates faster than the azimuthal
angle $\phi$, but since the total rotation of $\eta$ should be still $\pi$ when
$\phi$ is changing up to $2\pi$, as it is dictated by the winding number
$k=\half$, there will be three roots of the equation $\eta(\phi)=\phi$. These
roots, 0 and $\pm\phi_m$, define three straight polarization lines terminating
in the vortex center for the monstar geometry.\footnote{Note that for the
monstar all polarization lines residing within the sector $-\phi_m<\phi<\phi_m$
terminate in the vortex center, but only three of them are straight, i.e., are
having nonzero inclination at $r\rightarrow0$ (see \cite{Dennis09} for the
details).} One can see from Fig.~\ref{Fig1-Angles}a and \ref{Fig3-LemonStar}a
that when $\gamma$ approaches $-1$ the derivative becomes very close to unity,
but it never becomes bigger than 1, so that the monstar is not formed. The
reason preventing the appearance of the monstar is that it is not energetically
favorable to satisfy the condition \eqref{GM_MonstarCond}. In fact, it is the
most energetically favorable to have $\eta=\phi$, as for the hedgehog---the
vortex having the smallest energy when $\gamma\rightarrow-1$ (see
Fig.~\ref{Fig2-Es}). The rotation of polarization of the half-vortex is also
synchronous with the azimuth in rather wide sector, but the polarization never
overruns the azimuth. The monstar half-vortices, however, are expected to be
found in the exciton-polariton condensates out of equilibrium
\cite{Liew08,Borgh10,Keeling11}, where their appearance is not restricted by
energetics.

\section{Four-component exciton condensates}
 \label{sec:FourComp}
Excitons formed by an electron and a heavy hole in the semiconductor quantum
wells can be in four spin states \cite{IvchenkoBook}. The states with the total
spin projection $\pm1$ are optically active. These bright excitons are formed
by the heavy hole with the spin $+\threehalves$ and the the electron with the
spin $-\half$, or by the $-\threehalves$ hole and the $+\half$ electron. The
other two states are hidden from the observer and are usually referred to as
the dark excitons. The total momentum of these states is $\pm2$ and they are
formed either by the $+\threehalves$ hole and the $+\half$ electron, or by the
$-\threehalves$ hole and the $-\half$ electron.

The exciton-polaritons discussed in the previous sections are coupled states of
quantum-well excitons and microcavity photons. Only the bright excitons are
involved in this coupling, and the resulting condensates are two-component.
Since the frequency of a single exciton-polariton is shifted down with respect
to the single exciton frequency by a half of the Rabi frequency, \index{Rabi
frequency} the presence of dark excitons is irrelevant in this case provided
the exciton-photon coupling is strong enough. Contrary, when pure exciton
condensates are of interest, all four exciton spin states should be, in
general, taken into account. The formation of exciton condensates is possible
for cold indirect excitons in coupled quantum wells
\cite{Butov02,Hang95,Butov98,Voros05,Butov07}. The life-time of these excitons
is long enough, the excitons can travel coherently over long distances, and the
condensates can be formed in quasi-equilibrium conditions. The presence of
four-component exciton condensates has also been experimentally demonstrated
recently \cite{Leonard09}.

The indirect excitons are dipoles oriented along the growth axis of the
semiconductor structure, and their main interaction is spin-independent
dipole-dipole repulsion. The condensate state, however, is defined by weak
spin-dependent interactions arising from electron-electron, hole-hole, and
exciton-exciton exchanges \cite{Rubo11}. In what follows, I will assume the
signs of these interactions to be such that they favor the distribution of
excitons over all four spin states, populating both bright and dark components.
This state is similar to the linearly polarized two-component condensates
described above, but there is one important qualitative difference between
them. The exchange scattering of two excitons can result in transformation of
their spin states \cite{Ciuti98}. Namely, two bright excitons can turn into two
dark ones after collision and vice versa. These processes are described
microscopically by the Hamiltonian
\begin{equation}
 \label{FC-Hmix}
 \hat{\mathcal{H}}_\mathrm{mix}=W\left[
 \hat{\psi}_{+2}^\dag\hat{\psi}_{-2}^\dag
 \hat{\psi}_{+1}^{\vphantom{\dag}}\hat{\psi}_{-1}^{\vphantom{\dag}}
 +\hat{\psi}_{+1}^\dag\hat{\psi}_{-1}^\dag
 \hat{\psi}_{+2}^{\vphantom{\dag}}\hat{\psi}_{-2}^{\vphantom{\dag}}
 \right].
\end{equation}

In mean-field approximation, the creation $\hat{\psi}_{\sigma}^\dag$ and
annihilation $\hat{\psi}_{\sigma}$ exciton operators ($\sigma=\pm1,\pm2$) are
replaced by the order parameter components, $\psi_{\sigma}^*$ and
$\psi_{\sigma}$, respectively. The contribution of the resulting exciton-mixing
term $\mathcal{H}_\mathrm{mix}$ into the total energy of the exciton condensate
depends on the relative phases of the components. The term of this type is
absent in the two-component exciton-polariton case. Remarkably, the mixing of
excitons always leads to the decrease of the condensate energy, which is
achieved by fixing the proper interrelation between the phases. Denoting by
$\theta_\sigma$ the phase of $\psi_\sigma$, one can see that the following
relation holds within the order parameter manifold
\begin{equation}
 \label{FC_PhaseLock}
 \theta_{+2}+\theta_{-2}-\theta_{+1}-\theta_{-1}=
 \begin{cases}
 0 \pmod{2\pi}, & \text{if $W<0$,} \\
 \pi \pmod{2\pi}, & \text{if $W>0$.}
 \end{cases}
\end{equation}

The mixing term \eqref{FC-Hmix} additionally favors the formation of the
four-component exciton condensate with equal occupations of the components.
This fact can be seen from different perspective. $\mathcal{H}_\mathrm{mix}$
describes the transformation of \emph{pairs} of excitons, and, in the same way
as in the BCS theory of the superconductivity, %
\index{Bardeen-Cooper-Shrieffer (BCS) theory} this term leads to the pairing of
particles. This pairing leads to a decrease in the energy of the system and
results in appearance of the gap in the excitation spectrum for one excitation
branch. The other three excitation branches are Bogoliubov-like. %
\index{Bogoliubov dispersion} This follows from the fact that the phase locking
condition \eqref{FC_PhaseLock} leaves there angles to be undefined, so that
there are three Goldstone modes apart from the gaped mode induced by the
mixing.

The effect of applied Zeeman field \index{Zeeman field} on four-component
exciton condensate is expected to be very spectacular \cite{Rubo11}. The Zeeman
splitting is different for dark and bright excitons: the $g$-factor is given by
the sum of the electron and hole $g$-factors for the former, and by their
difference for the latter. When such a field is applied to the exciton
condensate its action is two-fold. On the one hand, it polarizes the bright and
dark components with different degrees of circular polarization, and thus
reduces the Zeeman energy of the condensate. On the other hand, the induced
imbalance in the occupation of the components increases the energy of the
mixing term \eqref{FC-Hmix} and suppresses the gap in the spectrum discussed
above. The interplay between these two effects can lead to the first-order
transitions from the four-component exciton condensate to the two- or the
one-component condensates. Note also that due to the presence of
$\mathcal{H}_\mathrm{mix}$ the system of equations defining the concentrations
of the components of the exciton condensate in the Zeeman field is nonlinear,
contrary to the case of two component exciton-polariton condensate (see
Eq.~\eqref{ZF_TwoEqs}).

Finally, it is important to note that vortices in the four-component exciton
condensate in the presence of the mixing of the component are composite: the
vorticity of one component should be accompanied by the vorticity of another
component to satisfy the the phase-locking condition \eqref{FC_PhaseLock}. As a
result one expect twelve elementary vortices. These are four polarization
vortices (two in each components), and the eight paired half-vortices in the
bright and dark components.

\section{Conclusions and perspectives}
 \label{sec:Concl}
The mean-field approximation provides simple and reliable method to
study the polarization properties and topological excitations of
exciton-polariton and exciton condensates that possess two and four
components of the order parameter, respectively. This includes, in
particular, the description of the polarization of the ground state and
elementary excitations of the condensates and their change in applied
Zeeman field, as well as the description of the texture of vortices and
vortex interactions.

The elementary topological excitations in two-component
exciton-polariton condensates are four half-vortices $(k,m)$ with
$k,m=\pm\half$, characterized by half-quantum changes of polarization
and phase angles. In the absence of
transverse-electric-transverse-magnetic (TE-TM) splitting of the lower
polariton band there is no coupling between the left half-vortices
(with $km<0$) and the right ones (with $km>0$), and one expects two
decoupled Berezinskii-Kosterlitz-Thouless (BKT) superfluid transitions
happening at the same temperature in the system. The TE-TM splitting
results in two qualitative effects. First, the cylindrical symmetry of
the half-vortex field is spontaneously broken that leads to warping of
the polarization field around a half-vortex and to deviation of the
streamlines of the supercurrent from the perfect circles. Secondly,
there appears long-range interactions between left and right
half-vortices. These interactions are particularly important in the
case of large longitudinal polariton mass $m_l$ when it favors the
formation of hedgehog (monopole) vortices (1,0) from the
$(\half,\half)$ and $(\half,-\half)$ half-vortices. The peculiarities
of the superfluid transition in this case and related features of
polarization textures of the exciton-polariton condensates are subjects
of further studies. In what concerns the geometry of the half-vortex
field it is shown that only two configurations, lemon and star, are
realized. The monstar configuration is not energetically favorable for
any value and sign of TE-TM splitting.

The essential feature of four-component exciton condensates is the
presence of mixing and related phase locking between dark and bright
excitons. One expects a nontrivial Zeeman-field effect resulting in a
discontinuous change of the polarization state of the condensate in
course of the first-order transition. The presence of composite
vortices in different components should lead to the formation of
interesting polarization patterns in driven exciton condensates that
provide an important topic for investigation, both experimental and
theoretical.

\section*{Acknowledgements}
This work was supported in part by DGAPA-UNAM under the project No.
IN112310 and by the EU FP7 IRSES project POLAPHEN.

%
%

%

\begin{thebibliography}{99.}

\bibitem{Kasprzak06} J. Kasprzak, M. Richard, S. Kundermann, A. Baas, P.
    Jeambrun, J. M. J. Keeling, F. M. Marchetti, M. H. Szyma\'nska, R.
    Andr\'e, J. L. Staehli, V. Savona, P. B. Littlewood, B. Deveaud, Le Si
    Dang, Nature \textbf{443}, 409 (2006)

\bibitem{Balili07} R. Balili, V. Hartwell, D. Snoke, L. Pfeiffer, and
    K. West, Science \textbf{316}, 1007 (2007)

\bibitem{Lai07} C. W. Lai, N. Y. Kim, S. Utsunomiya, G. Roumpos, H.
    Deng, M. D. Fraser, T. Byrnes, P. Recher, N. Kumada, T. Fujisawa,
    Y. Yamamoto, Nature \textbf{450}, 526 (2007)

\bibitem{Baumberg08} J. J. Baumberg, A. V. Kavokin, S. Christopoulos, A. J.
    D. Grundy, R. Butt\'e, G. Christmann, D. D. Solnyshkov, G. Malpuech, G.
    Baldassarri H\"oger von H\"ogersthal, E. Feltin, J.-F. Carlin, and N.
    Grandjean, Phys. Rev. Lett. \textbf{101}, 136409 (2008)

\bibitem{Wertz09} E. Wertz, L. Ferrier, D. D. Solnyshkov, P.
    Senellart, D. Bajoni, A. Miard, A. Lema\^{\i}tre, G.
    Malpuech, and J. Bloch, Appl. Phys. Lett. \textbf{95}, 051108 (2009)

\bibitem{Wouters07a} M. Wouters and I. Carusotto, Phys. Rev. Lett.
    \textbf{99}, 140402 (2007)

\bibitem{Wouters07b} M. Wouters, I. Carusotto, and C. Ciuti,
    Phys. Rev. B \textbf{77}, 115340 (2007)

\bibitem{Keeling08a} J. Keeling and N. G. Berloff, Phys. Rev. Lett.
    \textbf{100}, 250401 (2008)

\bibitem{Borgh10} M. O. Borgh, J. Keeling, and N. G. Berloff, Phys. Rev. B
    \textbf{81}, 235302 (2010)

\bibitem{Aleiner11} I. L. Aleiner, B. L. Altshuler, and Y. G. Rubo,
    Phys. Rev. B \textbf{85}, 121301(R) (2012)


\bibitem{Shelykh06} I. A. Shelykh, Y. G. Rubo, G. Malpuech, D. D.
    Solnyshkov, and A. Kavokin, Phys. Rev. Lett. \textbf{97}, 066402 (2006)

\bibitem{Ho98} T.-L. Ho, Phys. Rev. Lett. \textbf{81}, 742 (1998)

\bibitem{Ohmi98} T. Ohmi, K. Machida, J. Phys. Soc. of Japan \textbf{67}, 1822
    (1998)

\bibitem{Pelissetto02} A. Pelissetto and E. Vicari, Phys. Rep. \textbf{368},
    549 (2002)

\bibitem{Rubo06} Y. G. Rubo, A. V. Kavokin, and I. A. Shelykh, Phys. Lett. A
    \textbf{358}, 227 (2006)

\bibitem{Keeling08b} J. Keeling, Phys. Rev. B \textbf{78}, 205316
    (2008)

\bibitem{Larionov10} A. V. Larionov, V. D. Kulakovskii, S. H\"ofling, C.
    Schneider, L. Worschech, and A. Forchel,
    Phys. Rev. Lett. \textbf{105}, 256401 (2010)


\bibitem{Abrikos57} A. A. Abrikosov, Zh. Eksp. Teor. Fiz. \textbf{32}, 1442
    (1957) [Sov. Phys. JETP \textbf{5}, 1174 (1957)]

\bibitem{Berez72} V. L. Berezinskii, Sov. Phys. JETP \textbf{32}, 493 (1970);
    \textbf{34}, 610 (1972)

\bibitem{KT73} J. M. Kosterlitz and D. J. Thouless, J. Phys. C \textbf{6} 1181
    (1973)

\bibitem{Koster74} J. M. Kosterlitz, J. Phys. C \textbf{7}, 1046 (1974)

\bibitem{Rubo07} Y. G. Rubo, Phys. Rev. Lett. \textbf{99}, 106401 (2007)

\bibitem{VolovikBook} G. E. Volovik, \emph{The Universe
    in a Helium Droplet}, The International Series of Monographs on Physics, vol.\,117
    (Oxford University Press, 2003)

\bibitem{Lagoudakis08} K. G. Lagoudakis, M. Wouters, M. Richard, A. Baas, I.
    Carusotto, R. Andr\'e, Le Si Dang, B. Deveaud-Pl\'edran,
    Nature Phys. \textbf{4}, 706 (2008)

\bibitem{Lagoudakis09} K. G. Lagoudakis, T. Ostatnick\'y, A. V. Kavokin,
    Y. G. Rubo, R. Andr\'e, and B. Deveaud-Pl\'edran, Science \textbf{326}, 974
    (2009)

\bibitem{Flayac10} H. Flayac, I. A. Shelykh, D. D. Solnyshkov, and G. Malpuech,
    Phys. Rev. B \textbf{81}, 045318 (2010)

\bibitem{Miller10a} M. Toledo Solano and Y. G. Rubo, Phys. Rev. B
    \textbf{82}, 127301 (2010)

\bibitem{Rajaraman} R. Rajaraman, \textit{Solitons and Instantons}
    (North-Holland, Amsterdam, 1989), chap. 3.4

\bibitem{Liew07} T. C. H. Liew, A. V. Kavokin, and I. A. Shelykh, Phys. Rev. B
    \textbf{75} 241301 (2007)

\bibitem{ChaikinLub} P. M. Chaikin and T. C. Lubensky,
    \textit{Principles of condensed matter physics} (Cambridge
    University Press, Cambridge, England, 1995), chap. 9.3

\bibitem{Miller10b} M. Toledo Solano and Y. G. Rubo, J. Phys.:
    Conference Series \textbf{210}, 012024 (2010)

\bibitem{Panzarini99} G. Panzarini, L. C. Andreani, A. Armitage,
    D. Baxter, M. S. Skolnick, V. N. Astratov, J. S. Roberts, A. V. Kavokin,
    M. R. Vladimirova, and M. A. Kaliteevski,
    Phys. Rev. B \textbf{59}, 5082 (1999)

\bibitem{Miller12} M. Toledo Solano, M. E. Mora-Ramos, and Y. G. Rubo,
    unpublished

\bibitem{Nye99} J. F. Nye, \textit{Natural Focusing and Fine Structure of
    Light} (Institute of Physics Publishing, Bristol, 1999)

\bibitem{Soskin01} M. S. Soskin, M. V. Vasnetsov, Proress in Optics
    \textbf{42}, 219 (2001)

\bibitem{Dennis09} M. R. Dennis, K. O'Holleran, M. J. Padgett, Progress in
    Optics \textbf{53}, 293 (2009)

\bibitem{Berry77} M. V. Berry, J. H. Hannay, J. of Phys. A
    \textbf{10} 1809 (1977)

\bibitem{Nye83} J. F. Nye, Proc. R. Soc. London A \textbf{389}, 279 (1983)

\bibitem{Liew08} T. C. H. Liew, Y. G. Rubo, and A.V. Kavokin, Phys.
    Rev. Lett. \textbf{101}, 187401 (2008)

\bibitem{Keeling11} J. Keeling, N. G. Berloff, arXiv:1102.5302

\bibitem{IvchenkoBook} E. L. Ivchenko, \emph{Optical Spectroscopy of
    Semiconductor Nanostructures} (Alpha Science International, Harrow, UK, 2005)

\bibitem{Butov02} L. V. Butov, A. C. Gossard, and D. S. Chemla, Nature
    (London) \textbf{418}, 751 (2002)

\bibitem{Hang95} M. Hagn, A. Zrenner, G. B\"ohm, G. Weimann, Appl. Phys. Lett.
    \textbf{67}, 232 (1995)

\bibitem{Butov98} L. V. Butov and A. I. Filin, Phys. Rev. B \textbf{58}, 1980
    (1998)

\bibitem{Voros05} Z. V\"or\"os, R. Balili, D. W. Snoke, L. Pfeiffer, and K.
    West, Phys. Rev. Lett. \textbf{94}, 226401 (2005)

\bibitem{Butov07} L. V. Butov, J. Phys.: Condensed Matter \textbf{19}, 295202
    (2007)

\bibitem{Leonard09} J. R. Leonard, Y. Y. Kuznetsova, Sen Yang, L. V. Butov, T.
    Ostatnicky, A. Kavokin, and A. C. Gossard, Nano Lett. \textbf{9}, 4204 (2009)

\bibitem{Rubo11} Y. G. Rubo and A. V. Kavokin, Phys. Rev. B \textbf{84}, 045309
    (2011)

\bibitem{Ciuti98} C. Ciuti, V. Savona, C. Piermarocchi, A. Quattropani, and
    P. Schwendimann, Phys. Rev. B \textbf{58}, 7926 (1998)


\end{thebibliography}
%

\end{document}